\documentstyle[11pt]{article}
\newcounter{multieqs}

\begin{document}
\hfill {\Large DTP-MSU 93-01, April 1993}
\vspace{5cm}
\begin{center}
{\LARGE      CHARGED STRINGY BLACK HOLES WITH NON--ABELIAN HAIR}
\end{center}
\begin{center}
        {\bf E.E. Donets } and {\bf D.V.Gal'tsov}\\[0.2cm]
\begin{em}
         Department of Theoretical Physics, Physics Faculty,\\
         Moscow State University, 119899 Moscow, Russia\\
\end{em}
\end{center}

\vspace{2.5cm}

\begin{center}
{\bf Abstract}
\end{center}
Static spherically symmetric asymptotically flat charged
black hole solutions are constructed within the magnetic $SU(3)$ sector
of the 4-dimensional heterotic string effective action.
They possess non--abelian hair in addition to the
Coulomb magnetic field and are qualitatively similar to the
Einstein--Yang--Mills colored $SU(3)$ black holes except for the extremal case.
In the extremality limit the horizon shrinks and the resulting geometry around
the origin coincides with that of an extremal abelian dilatonic black hole with
magnetic charge. Non-abelian hair exibits then typical sphaleron structure.

\vfill \eject

Stringy four-dimensional charged black holes attracted much attention
recently in connection with the problem of the final stage of
Hawking evaporation \cite{ga}, \cite{gid}, \cite{wil} . A novel
feature of the $U(1)$ charged dilatonic black holes is the geometric
structure of an infinite throat in the extremal limit. In the string frame
the geometry is regular and this suggests some new (though still qualitative)
resolution of the information puzzle \cite{banks}. This family of solutions
is quite different from the supersymmetric $U^2(1)$ ones \cite{kal} possessing
both electric and magnetic charges and having the standard Reissner-Nordstrom
features in the extremal limit.

These results were obtained within the abelian Einstein--Maxwell--dilaton
model.  Within the context of some larger non--abelian gauge group, which is
associated with the string theory, similar solutions can be found as embedded
ones. In view of the above discussion any additional structures which can
emerge in the non--abelian case are of interest. In the absense of gauge
charges
the existence of a two-parametric family of essentially non-abelian dilatonic
black holes was shown recently \cite{dg}. From the other hand, in the pure
Einstein--Yang--Mills model (without dilaton) with the $SU(3)$ group
there exist charged black hole solutions possessing non-abelian hair
\cite{gv}. They have non--trivial extremal limit in which
non--abelian hair  survives. The $SU(3)$ group is the minimal one for which
non--linear superposition of a Coulomb field and a non-abelian hair is not
forbidden by the non-abelian baldness theorem \cite{ge}.

It can anticipated that analogous non-abelian charged black holes can exist in
the string theory with appropriate modifications due to the dilaton. We show
here that the bosonic effective action in the heterotic string theory admits
indeed static spherically--symmetric magnetically charged black holes
possessing non-abelian hair. They have, however, a different extremal limit as
compared with the no-dilaton case. Instead, this limit looks quite similar to
that of the abelian magnetically charged dilatonic black hole.
When dilaton is added, the radius of the horizon of an abelian extremal
solution
tends to zero. In this limit the dilaton field and the abelian component of
magnetic field both grow infinitely, while the non-abelian part remains finite.

We start with the field model which is the bosonic part of the 4-dimensional
heterotic string effective action in the Einstein frame without
axion field (for pure magnetic configurations the axion field is trivial,
see \cite{dg}) and without Gauss-Bonnet term
\begin{equation}
S = \frac{1}{16\pi} \, \int \, \sqrt{-g}\, [\, m^2_{Pl}\,(- R
+\, 2 \partial_{\mu} \Phi \partial^{\mu} \Phi) -
 \exp (-2 \Phi) \, F_{a\mu\nu} \, F_a^{\mu\nu}]\, d^4x \, ,
\end{equation} 
\noindent
where $\Phi$ is the dilaton, $F_{a \mu \nu}$ is the Yang-Mills
curvature corresponding to the $SU(3)$ gauge group.

Consider static spherically symmetric space-time with the line element
\begin{equation}
ds^2=\frac{\Delta \sigma^2}{r^2} dt^2 - \frac{r^2}{\Delta}dr^2
 -r^2 (d\theta^2 + \sin^2 \theta d\phi^2)\,.
\end{equation} 
\noindent
where $\Delta$ and $\sigma$ are functions of $r$.

Spherically--symmetric $SU(3)$ Yang--Mills connections may be classified
in terms of embeddings of the $SU(2)$ subgroup. For our purposes the
isospin-1 embedding is relevant (for more details see \cite{gv}) which
in the gauge $A_0=0$ reads
\begin{equation}
gA_i = T_a \epsilon_{aij} \frac{x^j}{r^2} (1-K)\, +\,
(\epsilon_{is\alpha} x_{\beta}\,+\,\epsilon_{is\beta} x_{\alpha})
\frac{x^s}{r^3} K_1\;.
\end{equation} 
\noindent
Here $T_a = \frac{1}{2} (\lambda_7, - \lambda_5,  \lambda_2)$
are normalized hermitean generators of the $SO(3)$ subgroup (with the standard
Gell-Mann matrices $\lambda_a$), $\alpha,\beta =1, 2, 3$ are matrix indices,
$g$ is the coupling constant, and $K, K_1$ are real--valued functions of the
radial variable. In this notation $x^j$ etc. are Cartesian coordinates related
to spherical ones in a standard flat--space way. In what follows we also
make an additional assumption for radial functions $K=K_1=f/{2\sqrt{2}}$ which
produces particular configurations with fixed value of a magnetic charge
\cite{gv}.

After dimensional reduction the action (1) will give
\begin{equation}
S = \frac{1}{2} \int dt dr \{m^2_Pl [\sigma'(\Delta/r - r)
- \Phi'^2 \Delta \sigma] - (2g)^{-2}\sigma F \exp (-2\Phi) \}\;,
\end{equation} 
\noindent
where
\begin{equation}
F=2\Delta f'^2 + f^4 -2 f^2 + 4
\end{equation} 
and primes denote the derivatives with respect to $r$.

The abelian (embedded) solution corresponds to $f\equiv 1$ and its geometry
reproduces the metric of Gibbons' magnetically charged dilatonic black hole
\cite{gi}. The particular value of the magnetic charge encountered here (as
well as in the no-dilaton version of the $SU(3)$ Einstein-Yang-Mills black
holes)
was pointed out by Marciano and Pagels long ago \cite{mp}. It is worse to be
remarked also that structure allowing for the gauge--invariant definition of
the magnetic charge (without Higgs scalars) emerges typically in the $SU(3)$
case \cite{mp} (for generalization to dyons see \cite{hp}. In the
pure Yang--Mills context corresponding solutions are point-like monopoles
(dyons) and they become black holes when gravity is added.
 From the other hand, the $SU(3)$ group is the one allowing for violation
of the non-abelian baldness theorem \cite{ge}. This opens the possibility
to consider non-abelian black holes possesing color charges. It should be
remarked the well-known $SU(2)$ non--abelian black holes have zero color charge
\cite{vg}; in this connection the name "colored" sometimes used for $SU(2)$
black holes with non-abelian hair \cite{bi} is somewhat misleading.

Variation of the action (4) gives the following set of equations
for $f, \Phi, \sigma$ and $\Delta$:
\begin{eqnarray}
(\frac{f' \Delta \sigma \exp(-2 \Phi)}{r^2})' =
\frac{\sigma f(f^2-1)\exp(-2\Phi)}{r^2} \; ,\\
(\Phi' \Delta \sigma)' = - \frac{R_g^2 \sigma F \exp(-2 \Phi)}
{4r^2}\;,\\
(\ln \sigma)' = \frac{R_g^2 f'^2 \exp(-2\Phi)}{2r} + r \Phi'^2\;,\\
- (\frac{\Delta}{r})' + 1 = \frac{R_g^2 F \exp(-2\Phi)}{4r^2} +
\Delta \Phi'^2 \;.
\end{eqnarray} 

We will be interested in the asymptotically flat configurations specified by
the asymptotic conditions $\sigma(\infty)=1$, $\Delta/r^2=1-2M/r+(P^2+D^2)/r^2
+O(1/r^3)$, where $O(1/r^2)$ terms are proportional to the sum of the magnetic
and the dilaton charges squared (in what follows we use this geometric
definition of the magnetic charge). Obviously, the Eq. 8 can be integrated
to give
\begin{equation}
\sigma=\exp[ - \int_r^{\infty} (\frac{R_g^2 f'^2 \exp(-2\Phi)}
{2r}+r \Phi'^2)dr]\, ,
\end{equation} 
\noindent
that can be used to reduce the system.  We also assume the existence on an
event horizon, $r=r_H$ being the largest root of $\Delta(r_H)=0$. The ADM mass
$M$ of the solution may then be expressed through the Eq.9 as
\begin{equation}
M=M_H+\frac{1}{2}\int_{r_H}^{\infty}(\Delta \Phi'^2 +
\frac{R^2_g F \exp(-2\Phi)}{4r^2})dr \,,
\end{equation} 
\noindent
where $M_H$ is a ``bare'' mass of a black hole.

For further simplification we fix the scale by imposing on the dilaton field
an asymptotic condition $\Phi (\infty) = 0$. Then using Eq.7 one can express
the dilaton charge of the configuration as follows
\begin{equation}
D= \lim_{r\rightarrow \infty}(-r^2 \Phi'(r)) = R_g^2 \int_{r_H}^{\infty} \frac
{\sigma F \exp(-2\Phi)}{4r^2} dr
\end{equation} 

An asymptotic behavior of the Yang-Mills function $f$ compatible with the
asymptotic flatness and corresponding to the magnetic charge $P=R_g\sqrt{3}/2$
is $f=\pm 1$. Then from the Eq. 6 it can be easily shown that
everywhere outside the horizon this function is bounded $f^2\leq 1$ \cite{ge}.
This property is used in the shooting strategy to obtain the solution
numerically. To implement this we first eliminate $\sigma$ through the Eq. 10
from the system (6)-(9) and then solve the remaining equations in terms of
power
series in the vicinity of the horizon. These expansions can be parameterized
by two quantities $\Phi_H$ and $f_H$ corresponding to (finite) values of the
Yang-Mills function and the dilaton field on the horizon
\begin{eqnarray}
f&=&f_H + \frac{f_H(f_H^2-1)}{2G_H x_H}\;y \,+\,O(y^2)\,,\\
\Phi&=&\Phi_H - \frac{y}{8x_H^2}(1-\frac{1}{G_H})  + O(y^2)\,,\\
\Delta&=& \frac{y}{2}\, R^2_g G_H\;(1+O(y))\,,
\end{eqnarray} 
\noindent
where $y=x-x_H$, $x=r^2/R_g^2$, and $G_H$ is the horizon value of the function
$G$ (in what follows we will use $R_g$ as the unit of length for all
parameters of the corresponding dimension such as $M$, $D$, and $P$)
\begin{equation}
G = 1 - \frac{[(1-f^2)^2+3]\exp(-2\Phi)}{4x}\,.
\end{equation} 

Eliminating from the Eqs.(6)-(9) the $\sigma$-variable through the Eq. 10
one gets in terms of dimensionless variable $x$ the following set of coupled
equations for three dimensionless quantities $f$, $\Phi$ and $d=\Delta/R_g^2$
\begin{eqnarray}
d (f_{xx} -2f_x \Phi_x) + \frac{1}{2}G f_x +
\frac{f(1-f^2)}{4x} = 0 \;,\\
d (\Phi_{xx} +\frac{1}{x} \Phi_x) +
\frac{1}{2} G \Phi_x + \frac{F\exp(-2\Phi)}{16x^2}=0 \;,\\
d_x  + d(2x\Phi_x^2 -\frac{1}{2x})+\frac{1}{2}
(\frac{F\exp(-2\Phi)}{x}-1)=0 \;.
\end{eqnarray} 
\noindent
The system consists of two equations of the
second order and one of the first order, hence the solution will be fixed
completely by the boundary conditions at the horizon for $f$, $f'$, $\Phi$,
$\Phi'$ and $d$, which are parameterized according to Eqs.
(13)-(15) in terms of $f_H$ and $\Phi_H$. The solution, like in the $SU(2)$
case \cite{dg}, exists for discrete values of the
parameters $f_H$ and $\Phi_H$, labeled by the number of zeros $n$ of
the Yang-Mills function $f$. For each integer $n$ and any (non-zero) real $x_H$
(dimensionless radius of the horizon) there exist a pair of values $f_H$ and
$\Phi_0$, and hence we obtain two-parametric family of solutions. These values
found numerically for some lower $n$ are shown on the Table 1 for $x_H=1$
together with the corresponding values of the $\sigma_H$, the
total mass, the dilaton charge (as given by Eqs. (11), (12)) and the Hawking
temperature measured in the units of $R_g^{-1}$
\begin{equation}
T=  \frac{\sigma (r_H)\,G_H}{4\pi x_H}\,.
\end{equation} 
\noindent
With increasing $n$ all these quantities are likely to tend to some limiting
values.

{\bf Table 1},  $x_H$ =1
\vskip5mm
\begin{tabular}{|c|c|c|c|c|c|c|}\hline
$n$ &$ f_H$  &$ \Phi_H$   &$ \sigma_H$ &$ M $  &$ D$      &$ T$       \\ \hline
0 & -1        & 0.458145 & 0.903508 & 0.790562 & 0.474341 & 0.050329\\
1 & -0.613879 & 0.522050 & 0.847105 & 0.858644 & 0.561263 & 0.047310\\
2 & -0.132473 & 0.548110 & 0.864465 & 0.865777 & 0.576706 & 0.046006\\
3 & -0.021835 & 0.549274 & 0.865984 & 0.866018 & 0.577332 & 0.045945\\
4 & -0.003561 & 0.549306 & 0.866025 & 0.866024 & 0.577348 & 0.045944\\ \hline
\end{tabular}
\vskip5mm
Here $n=0$ corresponds to an (embedded) abelian $f\equiv -1$ solution.
It can be seen that the value of the dilaton field on the horizon increases
with growing $n$, while the absolute value of the YM function $f$ is
decreasing.
The dilaton charge is substantially smaller than the Scwarzschild mass.

When the radius of the horizon decreases, the horizon values of the dilaton
field grow up and tend to the corresponding abelian value. At the same time
the horizon values of the YM function approach the abelian value $-1$, and
one can anticipate that ``small'' black holes in the vicinity of the horizon
look like their abelian counterparts. At larger distances the behavior of
the Yang--Mills function is qualitatively the same as previously (i.e.
oscillations around zero), and the mass of the solution increases with the
number of nodes. The dilaton charge becomes very close to the mass as in the
case of regular stringy sphalerons \cite{dg}. The Table 2 shows the parameters
of solutions up to $n=3$ for $x_H=0.0001$

{\bf Table 2}, $x_H=0.0001$
\vskip5mm
\begin{tabular}{|c|c|c|c|c|c|}\hline
$n$ &$ f_H$  &$ \Phi_H$   &$ \sigma_H$ &$ M$   &$ D$        \\ \hline
0 & -1        & 4.807898 & 0.016328 & 0.612393 & 0.612352 \\
1 & -0.999947 & 4.807956 & 0.016199 & 0.691746 & 0.695353 \\
2 & -0.999575 & 4.808081 & 0.016312 & 0.704072 & 0.703884 \\
3 & -0.997279 & 4.808844 & 0.016283 & 0.706391 & 0.706332 \\ \hline
\end{tabular}
\vskip5mm
Numerical solutions for $f$, $\Phi$, and $\sigma$ are shown at the Figs.1--4
(all functions at the Figs.1--4 depend on variable $r$; $r = R_g \sqrt{x}$
and we put $R_g = 1$).
The functions $\Phi(r)$ and $\sigma(r)$ are monotonic and rather similar
to those in the $SU(2)$ case \cite{dg}.

In the case of EYM (with no dilaton) $SU(3)$ black holes \cite{gv} there is
a critical (minimal) value of $x_H$ for which the coefficient in front of
the linear  in $x$ term in the expansion of the metric function $\Delta$ near
the horizon becomes zero, and consequently the leading term becomes quadratic
in $x$. This limiting value corresponds to the extremal charged black hole with
non-abelian hair and it marks a treshold horizon radius below which the family
of solutions ceases to exist. In the present case situation is rather
different.
As we have noticed, with decreasing $x_H$  the dilaton value on the horizon
is rapidly growing up, and consequently the second (negative) term in the
expression (15) is decreased with respect to the no-dilaton case. As a result,
the family of solution exists now for any arbitrarily small value of the radius
of the horizon.

The limiting form of solutions is of particular interest because of mentioned
above intriguing properties of the corresponding abelian solution. The $U(1)$
magnetically charged extremal black hole has $D=M$ , $|f|\equiv1$ and possess
the following expansions near the origin
\begin{eqnarray}
\Delta &=& \frac{r^2}{4} (1 + \frac{r^2}{2M^2}) + O(r^6),\\
\sigma &=& \frac{r}{M} (1 - \frac{r^2}{2M^2}) + O(r^5),\\
\exp(-2\Phi) &=& \frac{2r^2}{3} (1 - \frac{r^2}{2M^2}) + O(r^6).
\end{eqnarray} 
\noindent

In order to investigate the existence of larger family of extremal magnetically
charged black holes in the non-abelian case we look for somewhat more general
expansions of the type (21)-(23) near the origin. From the initial system of
equations (6)-(9) one can find the following family of approximate solutions
\begin{eqnarray}
\Delta &=& \frac{r^2}{4} (1 - k r^2) + O(r^6),\\
\sigma &=& {\sigma}_0\, r (1 + k r^2) + O(r^5),\\
\exp(-2\Phi) &=& \frac{2r^2}{3} (1 + k r^2) + O(r^6),\\
f &=& -1 +b r^2 +\frac{k-3b}{4} b r^4 + O(r^6),
\end{eqnarray} 
\noindent
where $k$, $\sigma_0$ and $b$ are some constant real parameters. Now we use
the same numerical strategy as above in order to match this expansions to the
required asymptotic form of the solutions. The matching procedure fixes the
discrete values of these parameters for each integer $n$, the number of nodes
of the Yang--Mills function $f$ . The Table 3 shows the results for $n=1,2,3$,
the abelian case $n=0$ being given for comparison.
\newpage
{\bf Table 3}
\vskip5mm
\begin{tabular}{|c|c|c|c|c|c|}\hline
$n$ &$ k$    & $b$       & $\sigma_0$ & $M$        & $D$        \\ \hline
0 & -1.33333 & 0         & 1.632993 & 0.612372 & 0.612372 \\
1 & -2.15396 & 1.076983  & 1.632998 & 0.687792 & 0.688132 \\
2 & -7.1271  & 8.502646  & 1.633002 & 0.703559 & 0.703916 \\
3 & -37.7495 & 54.43737  & 1.633014 & 0.706192 & 0.706561 \\ \hline
\end{tabular}
\vskip5mm

The abelian $n=0$ values of parameters correspond to the standard $U(1)$
extremal dilatonic black hole with the magnetic charge value
$P=\sqrt{3}/2$ as precribed by our embedding into $SU(3)$ in the adopted units.
The masses and dilaton charges are increased due to non--abelian hair. The
YM function oscillates as depicted on the Fig. 5 interpolating between the
value $-1$ at the horizon and $(-1)^{n+1}$ at infinity. The behavior of the
metric function $\sigma$ is not changed substantially as well as the dilaton
field for the number of nodes considered with respect to the abelian case,
see Figs. 6, 7.

In the string frame the corresponding geometry is that of an infinite throat
as can be seen from the expansion
\begin{equation}
ds_{string}^2 = \frac{3\sigma_0^2}{8} dt^2 - \frac{6}{r^2} dr^2 -\frac{3}{2}
(d\theta^2 +\sin^2\theta d\phi^2).
\end{equation}

This is the same kind of expressions as hold for the $U(1)$ extremal dilatonic
black holes which are just a particular case of (28) with $\sigma_0 = 1/M$.

We conclude with the following remarks. Non--abelian embedding of the $U(1)$
magnetic stringy black holes open the possibility of additional hair structure
similarly to the no-dilaton case. The role of dilaton, however, is essential
in the extremality limit, in which the radius of the horizon tends to zero
as in the abelian case. A novel feature of the non-abelian extremal
magnetically
charged dilatonic black holes is that the function $f$ now can interpolate
between the values $-1$ and $1$ (for odd $n$) corresponding to topologically
distinct Yang-Mills vacua. This property is quite similar to that of regular
sphaleron solutions in both the EYM \cite{gv1} and the EYM-dilaton \cite{dg}
theory.

\newpage
\begin{center}
{\Large Figure Captions}
\end{center}

Fig. 1. Yang--Mills field function $f$ for ``small'' black hole ($r_H=0.01$).

Fig. 2. Yang--Mills field function $f$ for ``medium size'' black hole
($r_H=1$).

Fig. 3. Dilaton field for $r_H=1, 0.01$ and $n=2$.

Fig. 4. Metric function $\sigma$ for $r_H=1, 0.01$ and $n=1, 3$.

Fig. 5. Yang--Mills function for the extremal solution, $n= 1,2,3$.

Fig. 6. Metric function $\sigma$ for extremal abelian $n=0$ and non--abelian
        $n=3$ solutions ($n=1,2$ curves being between the showed ones).

Fig. 7. Dilaton field for extremal abelian and non--abelian black holes.

\end{document}